# Spatial variations in aromatic hydrocarbon emission in a dust-rich galaxy

Justin S. Spilker[1], Kedar A. Phadke[2,3], Manuel Aravena[4], Melanie Archipley[2,3], Matthew B. Bayliss[5], Jack E. Birkin[1], Matthieu Béthermin[6], James Burgoyne[7], Jared Cathey[8], Scott C. Chapman[7,9,10], Håkon Dahle[11], Anthony H. Gonzalez[8], Gayathri Gururajan[6,12,13], Christopher C. Hayward[14], Yashar D. Hezaveh[15,14,16,17], Ryley Hill[7], Taylor A. Hutchison[18], Keunho J. Kim[5], Seonwoo Kim[2], David Law[19], Ronan Legin[15,14,16,17], Matthew A. Malkan[20], Daniel P. Marrone[21], Eric J. Murphy[22], Desika Narayanan[8,23,24], Alex Navarre[5], Grace M. Olivier[1], Jeffrey A. Rich[25], Jane R. Rigby[18], Cassie Reuter[2,3], James E. Rhoads[18], Keren Sharon[26], J.D. T. Smith[27], Manuel Solimano[4], Nikolaus Sulzenauer[28], Joaquin D. Vieira[2,3,29], Axel Weiß[28], Katherine E. Whitaker[30,24]

[1] Department of Physics and Astronomy and George P. and Cynthia Woods Mitchell Institute for Fundamental Physics and Astronomy, Texas A&M University, 4242 TAMU, College Station, TX 77843-4242, USA
[2] Department of Astronomy, University of Illinois, 1002 West Green St., Urbana, IL 61801, USA
[3] Center for AstroPhysical Surveys, National Center for Supercomputing Applications, 1205 West Clark St., Urbana, IL 61801, USA
[4] Núcleo de Astronomía de la Facultad de Ingeniería y Ciencias, Universidad Diego Portales, Av. Ejército Libertador 441, Santiago, Chile
[5] Department of Physics, University of Cincinnati, Cincinnati, OH 45221, USA
[6] Aix Marseille Univ., CNRS, CNES, LAM, Marseille, France
[7] Department of Physics and Astronomy, University of British Columbia, 6225 Agricultural Rd., Vancouver, V6T 1Z1, Canada
[8] Department of Astronomy, University of Florida, 211 Bryant Space Sciences Center, Gainesville, FL 32611, USA
[9] National Research Council, Herzberg Astronomy and Astrophysics, 5701 West Saanich Rd., Victoria, V9E 2E7, Canada
[10] Department of Physics and Atmospheric Science, Dalhousie University, Halifax, Nova Scotia, Canada
[11] Institute of Theoretical Astrophysics, University of Oslo, P.O. Box 1029, Blindern, NO-0315 Oslo, Norway
[12] Department of Physics and Astronomy 'Augusto Righi' (DIFA), University of Bologna, Via Gobetti 93/2, I-40129, Bologna, Italy
[13] INAF - Osservatorio di Astrofisica e Scienza dello Spazio, Via Gobetti 93/3, I-40129, Bologna, Italy
[14] Center for Computational Astrophysics, Flatiron Institute, 162 Fifth Avenue, New York, NY 10010, USA
[15] Département de Physique, Université de Montréal, Montréal, Québec, H3T 1J4, Canada
[16] Ciela - Montreal Institute for Astrophysical Data Analysis and Machine Learning, Montréal, Canada
[17] Mila - Québec Artificial Intelligence Institute, Montréal, Canada
[18] Observational Cosmology Lab, Code 665, NASA Goddard Space Flight Center, 8800 Greenbelt Rd., Greenbelt, MD 20771, USA
[19] Space Telescope Science Institute, 3700 San Martin Dr., Baltimore, MD 21218, USA
[20] Department of Physics and Astronomy, University of California, Los Angeles, CA 90095-1547, USA
[21] Steward Observatory, University of Arizona, 933 North Cherry Ave., Tucson, AZ 85721, USA
[22] National Radio Astronomy Observatory, 520 Edgemont Rd., Charlottesville, VA 22903, USA
[23] University of Florida Informatics Institute, 432 Newell Dr., CISE Bldg E251, Gainesville, FL 32611, USA
[24] Cosmic Dawn Center, DTU Space, Technical University of Denmark, Elektrovej 327, Kgs. Lyngby, DK-2800, Denmark
[25] The Observatories of the Carnegie Institution for Science, 813 Santa Barbara St., Pasadena, CA 91101, USA
[26] Department of Astronomy, University of Michigan, 1085 S. University Ave., Ann Arbor, MI 48109, USA
[27] Ritter Astrophysical Research Center, Department of Physics and Astronomy, University of Toledo, Toledo, OH 43606, USA
[28] Max-Planck-Institut für Radioastronomie, Auf dem Hügel 69, D-53121 Bonn, Germany
[29] Department of Physics, University of Illinois, 1110 West Green St., Urbana, IL 61801, USA
[30] Department of Astronomy, University of Massachusetts, Amherst, MA 01003, USA

**Dust grains absorb half of the radiation emitted by stars throughout the history of the universe, re-emitting this energy at infrared wavelengths.[1-3] Polycyclic aromatic hydrocarbons (PAHs) are large organic molecules that trace millimeter-size dust grains and regulate the cooling of the interstellar gas within galaxies.[4,5] Observations of PAH features in very distant galaxies have been difficult due to the limited sensitivity and wavelength coverage of previous infrared telescopes.[6,7] Here we present JWST observations that detect the 3.3μm PAH feature in a galaxy observed less than 1.5 billion years after the Big Bang. The high equivalent width of the PAH feature indicates that star formation, rather than black hole accretion, dominates the infrared emission throughout the galaxy. The light from PAH molecules, large dust grains, and stars and hot dust are spatially distinct from one another, leading to order-of-magnitude variations in the PAH equivalent width and the ratio of PAH to total infrared luminosity across the galaxy. The spatial variations we observe suggest either a physical offset between the PAHs and large dust grains or wide variations in the local ultraviolet radiation field. Our observations demonstrate that differences in the emission from PAH molecules and large dust grains are a complex result of localized processes within early galaxies.**

We observed the galaxy SPT0418-47 (SPT-S J041839-4751.9) as part of the Targeting Extremely Magnified Panchromatic Lensed Arcs and Their Extended Star formation (TEMPLATES) Early Release Science program on JWST. This galaxy was selected from the 2,500deg² South Pole Telescope catalog[8,9] of sources with dust-like spectra at millimeter wavelengths. Subsequent observations with the Atacama Large Millimeter/submillimeter Array (ALMA) confirmed SPT0418-47 as a z =4.2248 galaxy magnified by a factor of ~30-35 by strong gravitational lensing into a hallmark Einstein ring morphology by a foreground galaxy at z = 0.263.[10-12] Observations of the 158μm fine-structure line of ionized carbon suggested that this galaxy is isolated, with kinematics strongly dominated by rotational motions,[13] though a re-analysis of existing data and new JWST imaging[14] have identified a merging companion galaxy with a mass ratio of ≈1:5. JWST and additional submillimeter wavelength observations both suggest that the gas in SPT0418-47 has already been enriched in heavy elements to nearly the solar abundance.[15] After correction for the lensing magnification, the star formation rate of SPT0418-47 derived from its infrared (IR) luminosity is 350 ± 60 $M_\odot$/yr (here and elsewhere the error range quoted corresponds to a 1σ uncertainty), sufficiently high to drive a galaxy-scale outflow of cold molecular gas from the galaxy.[16]

We used JWST's Mid-InfraRed Instrument (MIRI) in its Medium Resolution Spectroscopy (MRS) mode to observe the redshifted 3.3μm PAH feature. The 3.3μm line, the shortest-wavelength PAH feature, arises due to carbon-hydrogen vibrational modes in the PAH molecules and is preferentially emitted only by the smallest neutral PAH molecules (less than ≈100 carbon atoms).[17,18] The MIRI/MRS operates as an integral field spectrograph, allowing the PAH emission to be spatially resolved. We processed the JWST data with the JWST pipeline software, including additional background subtraction and data artifact removal techniques (Methods).

Figure 1 shows the spectrum extracted from the MRS data cube over an annular aperture matched to the Einstein ring morphology seen at far-infrared wavelengths. Continuum emission arising from some combination of starlight and/or hot dust illuminated by an active galactic nucleus (AGN) is detected across the bandpass. We identify PAH emission at rest-frame 3.3μm, detected at integrated signal-to-noise ≈20. A possible second feature is detected at rest-frame 3.4μm, but the lack of spectral coverage redward of this feature makes its detection more tentative because the underlying continuum is poorly constrained. If real, this feature likely arises from PAHs with an aliphatic (i.e. lacking aromatic hydrocarbon rings) subgroup, although its origin is debated.[19-21] We see no evidence of absorption due to the 3.1μm water ice feature detected in 30% of low-redshift IR-luminous galaxies,[22] though the limited bandpass again precludes strong conclusions on this point. We compare to recent JWST/NIRSpec observations of the star-forming nucleus in the z = 0.02 galaxy merger VV114, which also shows strong 3.3μm PAH and 3.4μm aliphatic emission in addition to strong water absorption.[23]

As the first space-based mid-infrared integral field unit, the MIRI/MRS allows the PAH and underlying continuum emission to be spatially resolved at diffraction-limited 0.65" angular resolution. We collapsed the MRS data cube over the line-free spectral channels to create a map of the continuum emission, and used this image to subtract the continuum emission from the channels containing PAH emission (Methods). Figure 2 shows the continuum-subtracted 3.3μm PAH emission and the underlying rest-frame 3.1μm continuum emission. We also compare to archival ALMA imaging of the rest-frame 160μm dust continuum emission at matched spatial resolution (Methods). The spatial distribution of all three components is clearly distinct, an indication that the intrinsic lensing-corrected distribution of PAH, mid- and far-IR continuum emission are all different from one another. For the mid-IR and far-IR continuum, qualitatively similar morphologies are seen in MIRI F1800W imaging and ancillary ALMA data at similar spatial resolution, respectively,[12,14] lending further credence to the morphology we observe. While we do not model the lensing deflections or reconstruct the intrinsic source emission in this work, the typical physical scale probed in these comparisons is expected to be ~800pc due to the ~30x lensing magnification.

Extensive previous studies of low-redshift IR-luminous galaxies have demonstrated that the 3.3μm PAH equivalent width is a sensitive diagnostic of the presence of an obscured AGN. Hot dust surrounding the AGN dominates the mid-IR continuum, lowering the PAH equivalent width below a threshold value empirically found in the range 40-60nm by previous works.[22,24,25] Figure 3a shows a map of the 3.3μm PAH equivalent width created from the MRS data. We find equivalent width variations of a factor of 10 across the individual pixels of SPT0418-47, a reflection of the spatial offset between the PAH and continuum emission. Gravitational lensing effects cannot be responsible for these variations, because lensing has no wavelength dependence and we have ensured all comparisons are made at matched spatial resolution. Fig. 3 compares

SPT0418-47 to a sample of low-redshift IR-luminous galaxies detected in 3.3μm PAH emission by the Akari satellite[25] and five z ~ 2-3 galaxies with PAH detections from the Spitzer Space Telescope.[26,27] We make these comparisons using the surface density of IR emission $\Sigma_{IR}$, which is conserved by gravitational lensing.

On global scales, the measured 3.3μm PAH equivalent width 125±15nm is well above all thresholds for the presence of an obscured AGN, instead suggesting a system dominated solely by star formation. Even on resolved scales, Fig. 3b shows that very few pixels fall below the 40-60nm threshold values suggested as AGN diagnostics. It is plausible that this is a resolution effect: if the region influenced by the putative AGN is much smaller than the 0.65" MRS spatial resolution at this wavelength, the expected low PAH equivalent width could be diluted by the surrounding high equivalent width star-forming regions. Even in this scenario, however, it is clear that the emission from a hypothetical AGN has not yet come to dominate the galaxy; the JWST data clearly rule out a dominant obscured AGN in this source. This result is in contrast with the only other z ~ 4 galaxy detected in PAH emission,[7] where the low 6.2μm PAH equivalent width suggests the presence of a dust-obscured AGN.

The complex spatial distribution of PAHs in SPT0418-47 is also evident in comparison to the larger millimeter-size dust grains responsible for the far-IR continuum emission. Fig. 3c illustrates these variations in comparison to the ALMA rest-160μm continuum, which we take as a simple proxy for the total IR luminosity arising from large dust grains (an assumption we test in the Methods). The PAH emission in SPT0418-47 is weaker than all previous detections in galaxies at z > 1, but this is certainly a selection effect due to the low sensitivity of the Spitzer Space Telescope at mid-IR wavelengths. As with the equivalent width, we again observe factor-of-five variations in the $L_{PAH}/L_{IR}$ ratio, driven by the spatial mismatch between the PAH and far-IR emission. On resolved scales, the individual pixels in the distribution of $L_{PAH}/L_{IR}$ trace out a similar range as that seen in the galaxy-averaged low-redshift sample, with the PAH emission becoming progressively weaker towards high $\Sigma_{IR}$. This suggests that the deficit of PAH emission arises on local scales within individual galaxies, and is not merely reflective of scatter in the overall galaxy population.

The origin of the spatial variations in 3.3μm PAH emission and the larger dust grains responsible for the far-IR emission is unclear. We have made no correction for possible extinction at 3.3μm, which may be relevant if the PAH molecules are well-mixed with the larger dust grains in locations with equivalent visible wavelength attenuation $A_V$ ~ 30mag.[25,28,29] Arising from the smallest PAH molecules, the 3.3μm feature is especially sensitive to the local UV radiation field, dust destruction processes such as shocks or photodissociation, and grain growth processes that enhance PAH features at longer wavelengths to the detriment of the 3.3μm feature.[17] The spatial variations may thus reflect a highly non-uniform UV radiation field or genuine differences in the distribution of the PAH molecules and millimeter-size dust grains. Whatever the cause, it is clear that the 3.3μm PAH feature is a poor tracer of the total IR luminosity on sub-galactic scales,

calling into question its use as a direct indicator of the star formation rate of high-redshift galaxies.[24]

We have detected and resolved the 3.3μm PAH emission in a merger-driven star-forming galaxy at z = 4.2248, the most distant detection of any complex aromatic molecule so far. Our results demonstrate spatial differences in the 3.3μm PAH, 3.1μm continuum, and 160μm continuum emission, suggesting either that the small and large dust grains are not co-located within the galaxy or the presence of large variations in the heating of the dust grains by the local UV radiation field. Whether such spatial offsets are a common feature of high-redshift galaxies must be further investigated by future JWST observing campaigns now made possible by the high sensitivity and angular resolution of the MIRI/MRS instrument.

**Acknowledgements:** JWST is operated by the Space Telescope Science Institute under management of the Association of Universities for Research in Astronomy, Inc., under NASA contract NAS 5-03127. ALMA is a partnership of ESO (representing its member states), NSF (USA) and NINS (Japan), together with NRC (Canada), MOST and ASIAA (Taiwan), and KASI (Republic of Korea), in cooperation with

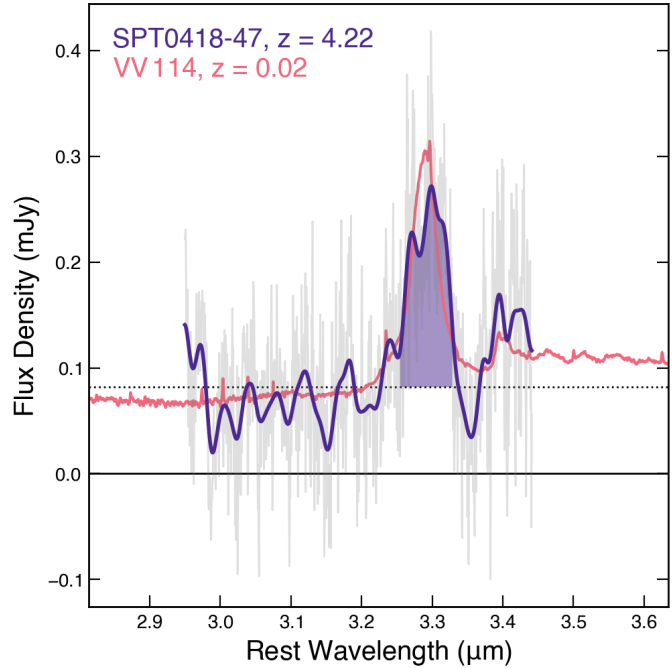

**Figure 1 | PAH 3.3μm emission from SPT0418-47.** The MRS spectrum is shown at both the native spectral resolution (thin gray line) and after smoothing to R ≈ 600 (navy). As a comparison, the JWST/NIRSpec spectrum of the star-forming nucleus in the low-redshift merger VV114 is shown in magenta,[23] re-scaled in amplitude to match the MRS spectrum at rest-frame 3.2μm. The dotted line shows the estimated continuum level, and the shaded channels illustrate the wavelength range used to create the integrated map of the PAH emission in Figure 2.

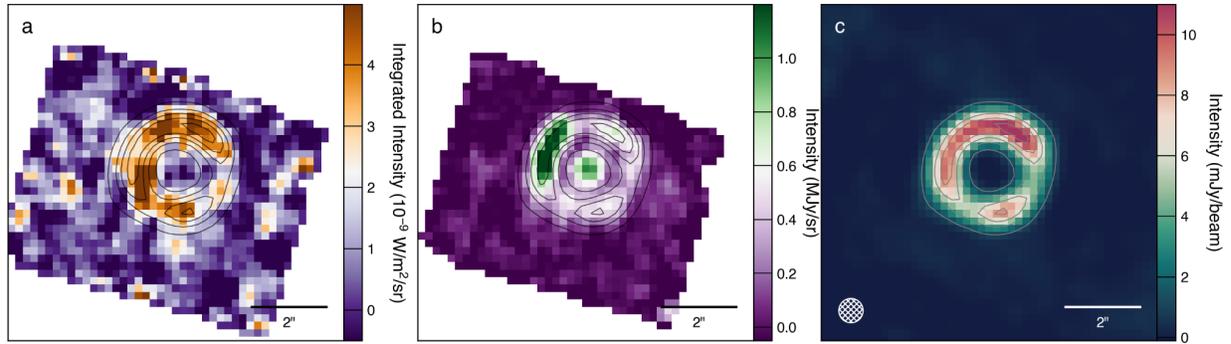

**Figure 2 | Mid- and far-infrared continuum and PAH emission from SPT0418-47. a**, Continuum-subtracted 3.3μm PAH emission integrated over the wavelength range indicated in Fig. 1. The range in integrated intensity is provided in the colorbar. **b**, Rest-frame 3.1μm continuum image created from the PAH-free channels of the MIRI/MRS data. The range in intensity is shown at right. Stellar emission from the foreground lens galaxy is visible at the center of the lensed Einstein ring. **c**, Rest-frame 160μm continuum image from ALMA matched to the 0.65" spatial resolution of the MIRI/MRS data (Methods). The color scale at right shows the flux density per synthesized beam, the shape of which is shown at lower left. Contours of the 160μm flux density are repeated in panels a-c at levels of 10, 30, 50, 70, and 90 times the noise level of 0.12mJy per beam.

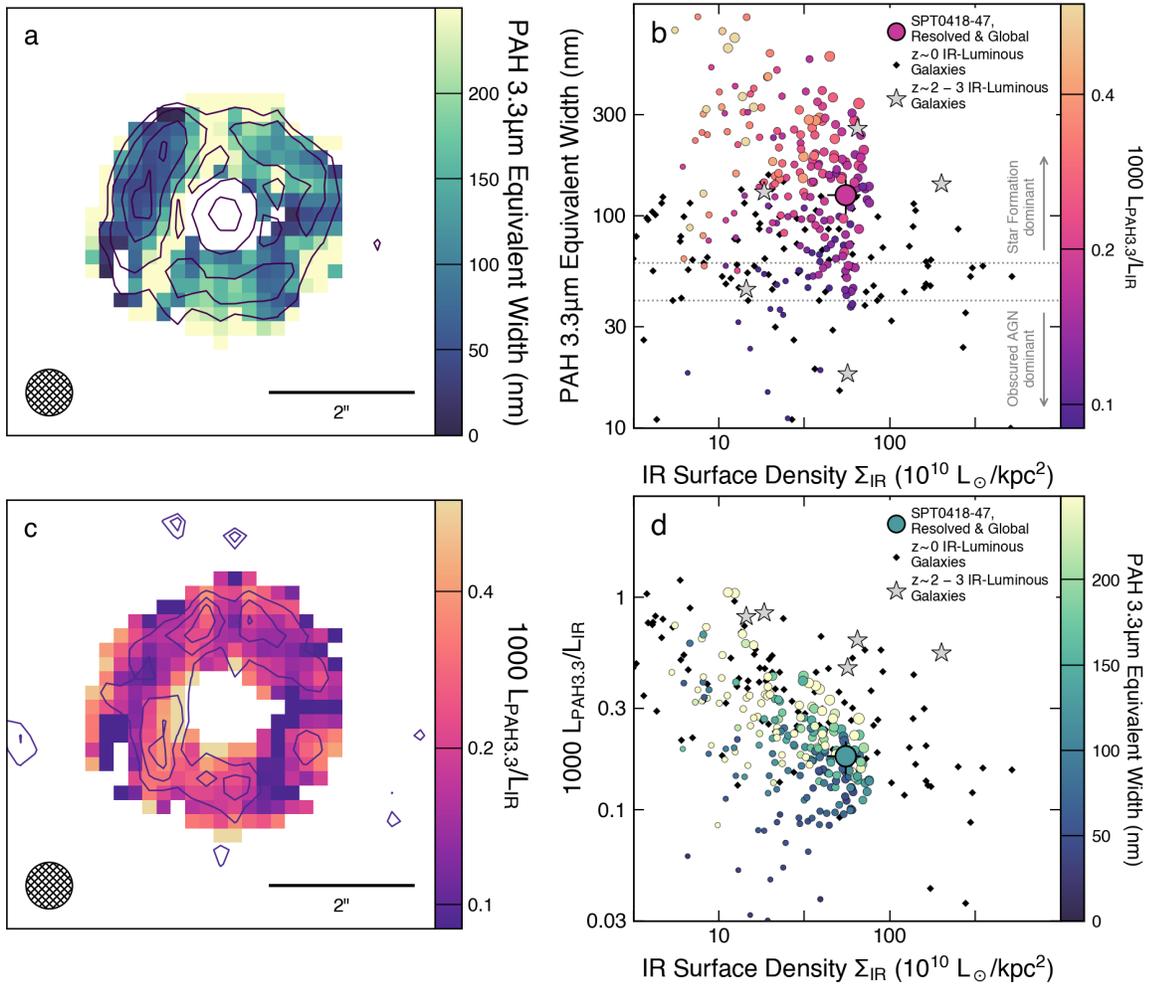

**Figure 3** | **Large spatial variations in PAH emission within SPT0418-47. a**, The 3.3μm PAH equivalent widths are high throughout SPT0418-47, but large variations are apparent as a result of the spatial offsets between PAH and underlying continuum. The image shows a map of the 3.3μm PAH equivalent width, the range of which is indicated at right. Dark contours show the rest-frame 3.1μm continuum emission, as in Fig. 2, and the ellipse at lower left indicates the effective spatial resolution of the data. **b**, On both spatially-integrated and resolved scales, the 3.3μm PAH equivalent width as a function of the IR surface density $\Sigma_{IR}$ demonstrates that an obscured AGN is not dominant within SPT0418-47. Individual pixels from (a) are shown as circles colored by their $L_{PAH}/L_{IR}$ ratio and sized in proportion to their 3.3μm PAH flux. Dotted horizontal lines indicate regimes in which either star formation (high equivalent width) or obscured AGN (low) are thought to dominate. **c**, The mismatch between emission from PAHs and larger dust grains is also evident in a map of the $L_{PAH}/L_{IR}$ ratio. Thick navy contours illustrate the PAH emission as in Fig. 2. **d**, Individual pixels within SPT0418-47 trace out a deficit of PAH emission with a similar distribution in $L_{PAH}/L_{IR}$ vs. $\Sigma_{IR}$ as the bulk of source-averaged low-redshift galaxies. Individual pixels from panel (c) are shown as circles colored by their 3.3μm PAH equivalent width and sized in proportion to their 3.3μm PAH flux. In panels (b) and (d) we compare to spatially-integrated measurements of both low-redshift galaxies[25] (black diamonds) and five galaxies with 3.3μm PAH detections from Spitzer[26,27] (grey stars).

## METHODS

**Cosmology and Initial Mass Function Assumptions.** Throughout this paper we assume a concordance flat ΛCDM cosmology with $\Omega_m$ = 0.3 and h = 0.7. We assume a Chabrier[30] initial mass function to convert between $L_{IR}$ and star formation rate. We convert all literature measurements to match these assumptions where necessary.

**JWST observations and analysis.** The TEMPLATES ERS program (#1355, co-PIs J. R. Rigby and J. D. Vieira) was primarily designed to study star formation indicators in distant galaxies while exercising the spectroscopic capabilities of JWST using both the NIRSpec[31,32] and MIRI/MRS[33-35] integral field observing modes. In this work we focus on MRS observations of SPT0418-47 targeting the 3.3μm PAH feature redshifted to 17.2μm, covered by Channel 3 of the LONG ('C') grating. The field-of-view is 5.2" x 6.2" and the spectral resolving power is ≈2600 at the wavelength of the redshifted PAH feature. We used the SLOWR1 detector readout pattern, a standard 4-point dither pattern, and achieved an on-source integration time of 3440s. The science observations were accompanied by dedicated background exposures using the same parameters.

The data were taken on 8 August 2022, and we re-reduced the raw frames using version 1.8.4 of the JWST pipeline and pipeline calibration context `jwst_1019.pmap`. This version of the pipeline contains a preliminary treatment of the so-called 'shower' artifacts, the origin of which is still under investigation but likely related to cosmic rays. The uncalibrated science and background frames were processed with the `Detector1` pipeline to apply detector-level corrections, fit the ramps, and identify and exclude cosmic ray and shower artifacts. The output rate images were processed outside the pipeline to flag additional bad pixels newly-identified using the dedicated background exposures. The `Spec2` pipeline applies distortion and wavelength calibration, flux calibration, and other corrections to the science frames.

We use 2-D pixel-by-pixel background subtraction in this stage, where the individual background exposures were median combined to create a composite background image subtracted from the science data. While this is noisier in principle than the use of a 1-D master background spectrum, it negates detector and flat-field systematics that would otherwise limit our ability to detect faint extended structures in the science data at long wavelengths. Finally the `Spec3` pipeline creates a 3-D data cube for each of the four channels observed in the LONG grating. The Channel 3 cube has 0.20" pixel scale in the spatial dimensions and 3nm channels in the wavelength dimension.

We extracted a spectrum of SPT0418-47 from the background-subtracted data cube using an annular aperture with inner and outer radii 0.6" and 1.8" centered on the source. We created a simple constant continuum image of the source by collapsing all channels of the cube, excluding the PAH feature. We neglect any possible contribution from the hydrogen Pf-δ 3.29μm line. This line is expected to be 10x fainter than the 4.05μm Br-α line, which is also not detected in our MRS data. While the continuum in this region of the spectrum is known to be complex,[29] the limited bandwidth of the cube precludes a more detailed treatment. We created a continuum-subtracted cube by subtracting this continuum image from each channel, and then created an image of only the PAH emission by summing over channels corresponding to rest-frame wavelengths 3.25-3.33μm. This range is justified by the expected intrinsic width of the 3.3μm PAH solid-state feature.[17,29]

**MRS residual artifact removal.** Although the JWST pipeline version we used has an early treatment to identify and flag cosmic ray shower artifacts, these artifacts are by far the dominant remaining instrumental artifact after pipeline processing and limit the sensitivity of our observations. Unflagged shower pixels are visible in the fully-calibrated 2-D frames and mapped into the 3-D cube. The regions on the detector affected by showers are typically several hundred pixels in area, with no obvious spatial structure. Due to the geometry of the slicing optics, these regions are generally contained within a single slice, but span tens of pixels in the along-slice and dispersion directions. When mapped from the 2-D frames into the 3-D data cube, the shower artifacts then manifest as stripes mostly aligned with the cube's x-dimension when the cube is built with the pipeline's `IFUALIGN` cube-building option. The artifacts are coherent over tens of channels in the wavelength dimension. Given the faintness of the PAH emission, these artifacts make subsequent analysis challenging. Moreover, because SPT0418-47 is an extended source and the PAH feature is intrinsically broad, true emission from the source would also exhibit similarly extended features in the 2-D frames, making a 2-D treatment of the background especially fraught. Instead we developed a background subtraction technique that largely removes these stripe artifacts in the 3-D cube, where the location of true source emission is well known. Presumably the removal of these shower artifacts will be an active area of research over the lifetime of JWST.

We used the `photutils` package to create an estimate of the residual stripe artifacts in each channel of the MRS data cube. We masked the location of the SPT0418-47 Einstein ring and foreground lens galaxy in every channel based on its known position and size from ALMA submillimeter observations to prevent true source emission from contaminating the background estimate. Because the stripes are coherent over tens of wavelength channels, we fit for the background using a 25 channel running average around each channel of the cube. The stripes are generally (though not perfectly, due to the curved slices of the IFU on the detector) aligned with the cube's x-dimension, so we force the background to consist of a series of rows in the x-direction, allowing a linear slope in the background from the left to the right half of the row. Each row was a single pixel in height, but we also tested variants using two- or three-pixel stripes. The two-pixel stripe template worked equally well as our one-pixel default, but the three-pixel variant still left notable striping artifacts, likely because the three-pixel version combines data from multiple slices from different regions of the detector with different shower structure.

Extended Data Figure 1 shows the results of this stripe removal process, arbitrarily averaging over 100-channel wavelength ranges in the data cube as a simple demonstration of the method. Recall that the background is estimated for each channel of the data cube using a 25 channel running average over the wavelength dimension. The stripes are (partly by construction) also visible in the estimated background over even larger wavelength intervals. The early JWST pipeline shower removal techniques alone leave significant artifacts that become visible due to the large spatial extent of the showers in the detector's y-dimension (cube wavelength dimension). The stripes that were removed well enough to be invisible in individual channels are now apparent when averaged over wavelength. The final result of this procedure yields a data cube with no stripe artifacts even when averaged in the wavelength dimension.

We have publicly released our full analysis scripts, including our custom artifact removal technique, on the TEMPLATES collaboration github repository <https://github.com/jwst-templates>.

**ALMA observations and analysis.** SPT0418-47 has been observed extensively by ALMA at many wavelengths. In this work we use the rest-frame 160μm observations obtained in projects 2016.1.01374.S (PI: Y. Hezaveh) and 2016.1.01499.S (PI: K. Litke); these data also covered the [CII] 158μm emission line. The data were reduced using standard pipeline scripts, and we performed a single round of phase-only self-calibration for each execution of the observing blocks. We jointly imaged the data using natural weighting of the visibilities, accounting for the primary beam response differences due to the different phase centers between the two ALMA projects.

Because we are primarily interested in the comparison between the PAH emission and the mid- and far-IR continuum, we used a combination of a *uv*-plane taper and the `restoringbeam` parameter of the CASA `tclean` task to produce a synthesized beam with 0.65" full-width-at-half-maximum and 0.2" pixel scale, matched to the diffraction-limited resolution of MIRI/MRS at the 17.2μm observed wavelength of the PAH feature. The synthesized beam is Gaussian by construction, while the MRS point spread function shows the characteristic star pattern of all JWST instruments. The S/N of the PAH detection is sufficiently low that the PSF wings have little influence on our comparison. The final ALMA 160μm image reaches a sensitivity of 120μJy/beam and detects the source at a peak significance of ≈110. This is far higher than the S/N achieved by the MRS data, so all statistical uncertainties in the cross-observatory comparison are solely dominated by the JWST rather than ALMA data.

We assume a simple linear conversion between the 8-1000μm IR luminosity and the 160μm continuum flux density. We take $L_{IR}$ from ref. [36], who fit the IR spectral energy distribution with a standard modified blackbody function and mid-IR power law, with photometric coverage from 9 bands spanning 100-3000μm (rest-frame 19-580μm). By assuming a constant conversion between $L_{IR}$ and 160μm continuum, we implicitly ignore any variations in the cold dust temperature across the source. Based on the distribution of the ratio of 160μm flux density to $L_{IR}$ across the SPT sample,[11,36] we estimate uncertainties of a factor of ≈1.8 on the IR luminosity of individual pixels in the resolved map.

We test our assumption of a constant conversion factor between $L_{IR}$ and 160μm continuum flux density using additional ALMA continuum data at rest-frame 120μm.[12] Variations in the cold dust temperature $T_{dust}$ will appear as changes in the 120μm/160μm flux ratio across the source, with higher $T_{dust}$ corresponding to larger ratios. We re-image the rest-120μm data following the same procedure as the 160μm data to produce an image with identical spatial resolution and pixel scale. We convert the 120μm/160μm flux ratio to an equivalent estimate of $T_{dust}$ by assuming that the dust continuum emission follows a standard modified blackbody function with Rayleigh-Jeans slope index $\beta = 2$ and unity optical depth wavelength $\lambda_0 = 100$μm. These images are shown in Extended Data Figure 2. We see no evidence that the regions of brightest emission are also preferentially warmer than the rest of the source. The largest variations arise in the faintest pixels, suggesting that instrumental noise is plausibly responsible for the variations we see.

At least at the spatial resolution of the MRS data, we find only small variations in $T_{dust}$ across the source. We further examine the implications of the $T_{dust}$ variations by calculating a 'corrected' conversion factor between $L_{IR}$ and 160μm flux density. Pixels with high 120μm/160μm flux ratios (high $T_{dust}$) will have higher $L_{IR}$ for a given amount of 160μm flux density. Extended Data Figure 2 (right) shows the results of this test in comparison to our default assumption of a constant conversion factor. In agreement with the relatively small implied variations in dust temperature from the ALMA maps, we find only very small ~10% corrections to our default assumption. These changes are minor in comparison to other sources of uncertainty in our analysis such as the relatively low S/N in the 3.3μm PAH data. While future work will examine the spatially-resolved dust emission from SPT0418-47 in more detail, we presently have no evidence for large variations in $T_{dust}$ that would lead to sharply skewed changes in $L_{IR}$ on resolved scales.

**PAH ratio recovery tests.** The 3.3μm PAH emission is relatively weak, and the MRS data cube shows structured noise even after all processing. It is possible these two factors conspire to produce an artificial distribution in, for example, the $L_{PAH}/L_{IR}$ ratio map (Fig. 3). We tested this scenario by constructing mock MRS cubes in which $L_{PAH}$ perfectly traces $L_{IR}$ on resolved scales, i.e. assuming a constant $L_{PAH}/L_{IR}$ ratio across the source. We rescaled the high signal-to-noise

ALMA 160μm image to match the integrated luminosity of the PAH feature in the real data, assuming a simple Gaussian profile in the wavelength direction with 4500km/s full width at half-maximum. We injected this artificial signal into the continuum-subtracted MRS cube, avoiding wavelengths near the real PAH feature. After continuum subtraction, the MRS cube contains essentially only noise, allowing the noise properties of the real data to influence the recovered $L_{PAH}/L_{IR}$ map. We perform several realizations of this exercise by varying where within the MRS cube the model signal is injected.

Extended Data Figure 3 demonstrates the results of this test. Even in the faintest pixels, the depth of the MRS data is sufficient to recover the true $L_{PAH}/L_{IR}$ ratio to within a factor of ≈2, improving to a factor of 25% in brighter regions. These variations are far smaller than the factor-of-10 variations in $L_{PAH}/L_{IR}$ we find in the real data, confirming that the source exhibits large variations in $L_{PAH}/L_{IR}$.

**Methods References.**

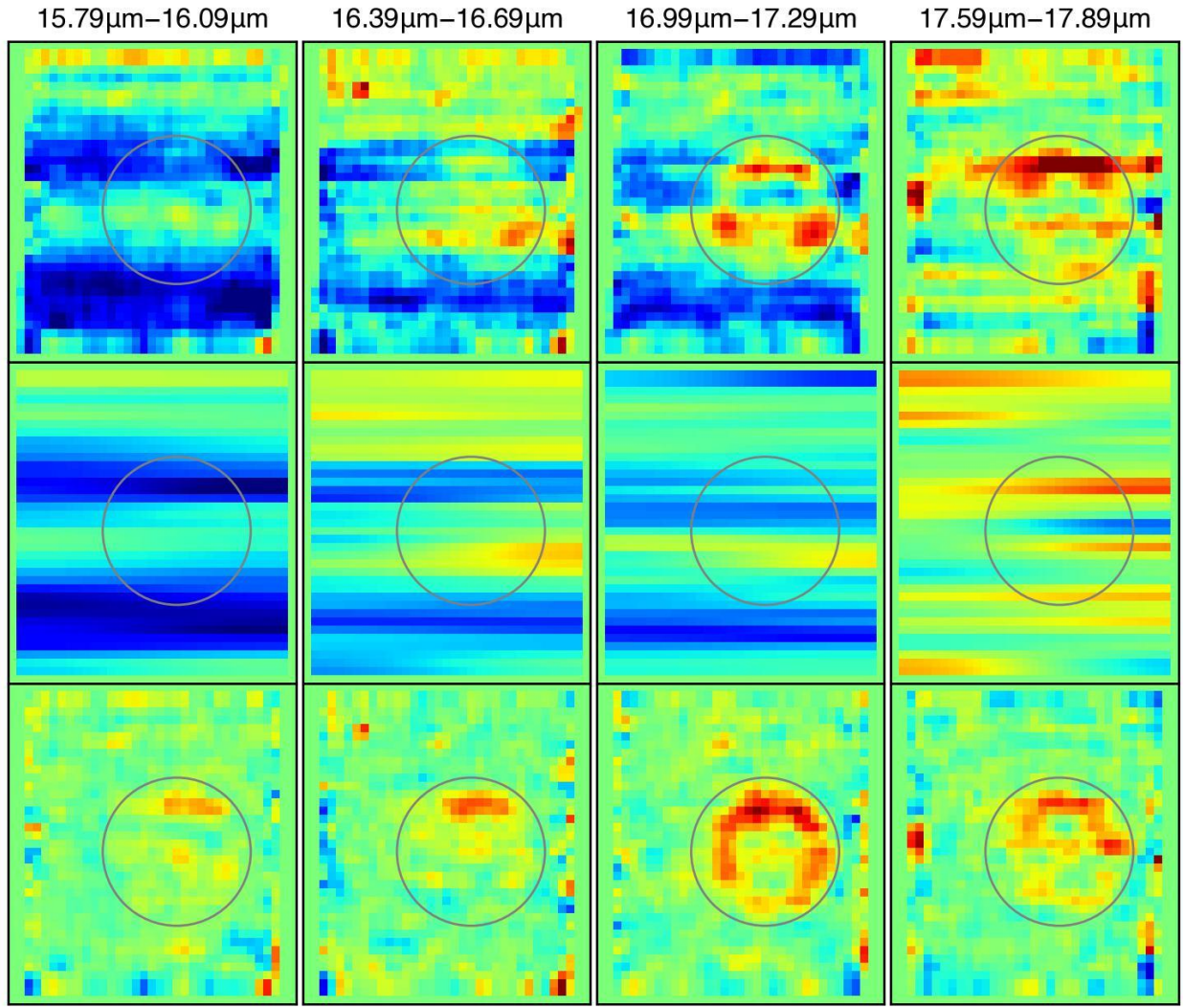

**Extended Data Figure 1 | Demonstration of residual MRS data artifact removal process.** Each column shows a 100-channel average from the MRS data cube corresponding to the wavelength ranges indicated at top. The top row shows the original pipeline-processed data. Horizontal stripe features are evident, a manifestation of the so-called 'shower' artifacts in MIRI data. The middle row shows the estimated background to be subtracted averaged over the same wavelength range. The bottom row shows the final background-subtracted image. The circles show the region of the cube that was masked during the background estimation due to the presence of real source emission from SPT0418-47. All images are on the same color scale. The 3.3μm PAH feature is mostly contained within the wavelength range of the third column.

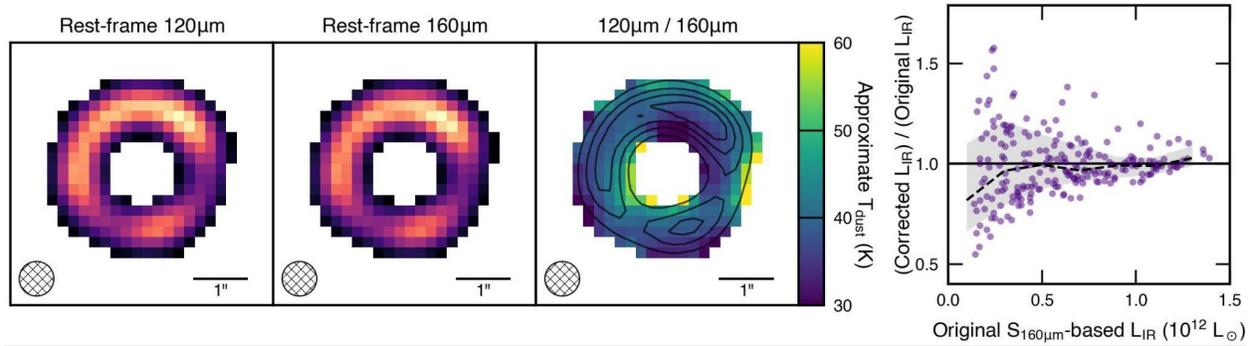

**Extended Data Figure 2 | Investigations of possible variations in dust temperature across SPT0418-47.** Using additional ALMA continuum data at rest-frame 120μm, we calculate the implied changes in $T_{dust}$ under standard assumptions about the shape of the dust spectral energy distribution. The 120μm and 160μm images are shown on a linear min/max color scale, masking pixels detected at S/N<5 in either band, to demonstrate their qualitative similarity. This similarity consequently implies only small changes in $T_{dust}$ across the source. We use the resolved $T_{dust}$ map to estimate the implied correction to our 'default' assumption of a constant conversion between $L_{IR}$ and 160μm flux density; the right panel shows that only ~10% variations are implied, subdominant to other sources of uncertainty in our analysis.

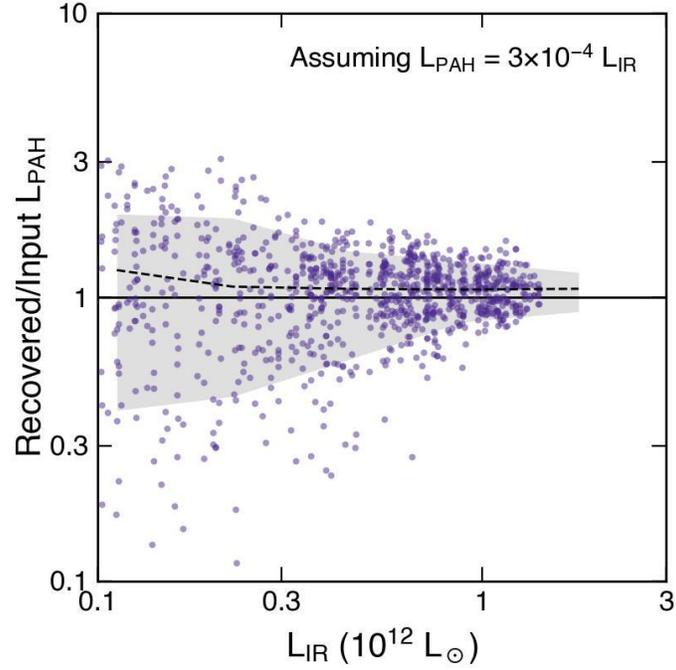

**Extended Data Figure 3 | Recovery of variations in $L_{PAH}/L_{IR}$.** Using mock data with constant $L_{PAH}/L_{IR}$ inserted into signal-free portions of the MRS data cube, we test the extent to which to faintness of the PAH feature and noise properties of the MRS data influence our conclusion that SPT0418-47 shows large variations in $L_{PAH}/L_{IR}$. Points show individual pixels from several of the mock realizations, while the black dashed line and grey shaded region illustrate the median and 16-84th percentile range of the distribution of all mock simulations. Even in the faintest regions, $L_{PAH}/L_{IR}$ is still recovered to within a factor of ≈2, improving to ±25% in brighter regions.